\documentclass[prb,twocolumn,showpacs,superscriptaddress,preprintnumbers,amsmath,amssymb]{revtex4}
\usepackage{graphicx}
\usepackage{dcolumn}
\usepackage{bm}

\def\G{\Gamma}
\def\z{\zeta}

\newcommand{\be}{\begin{eqnarray}}
\newcommand{\ee}{\end{eqnarray}}

\def\beqn{\begin{eqnarray}}
\def\eeqn{\end{eqnarray}}
\def\beq{\begin{equation}}
\def\eeq{\end{equation}}
\begin{document}
\title{Ordering near the percolation threshold in models of 2D interacting bosons with quenched dilution}
\author{N. Bray-Ali}
\affiliation{Department of Physics, University of California,
Berkeley, CA 94720}
\author{J.~E.~Moore}
\affiliation{Department of Physics, University of California,
Berkeley, CA 94720} \affiliation{Materials Sciences Division,
Lawrence Berkeley National Laboratory, Berkeley, CA 94720}
\author{T. Senthil}
\affiliation{Department of Physics, Massachusetts Institute of
Technology, Cambridge MA 02139}
\author{A.~Vishwanath}
\affiliation{Department of Physics, University of California,
Berkeley, CA 94720} \affiliation{Materials Sciences Division,
Lawrence Berkeley National Laboratory, Berkeley, CA 94720}
\date{July 25, 2005}
\pacs{05.30.Jp,64.60.Ak, 74.81.-g,75.10.Jm, 75.10.Nr, 75.40.Cx, 75.40.Mg}
 
\begin{abstract}
Randomly diluted quantum boson and spin models in two dimensions
combine the physics of classical percolation with the well-known
dimensionality dependence of ordering in quantum lattice models.  This
combination is rather subtle for models that order in two dimensions
but have no true order in one dimension, as the percolation cluster
near threshold is a fractal of dimension between 1 and 2: two
experimentally relevant examples are the $O(2)$ quantum rotor and the
Heisenberg antiferromagnet.  We study two analytic descriptions of the
$O(2)$ quantum rotor near the percolation threshold.  First a
spin-wave expansion is shown to predict long-ranged order, but there
are statistically rare points on the cluster that violate the standard
assumptions of spin-wave theory.  A real-space renormalization group
(RSRG) approach is then used to understand how these rare points
modify ordering of the $O(2)$ rotor.  A new class of fixed points of
the RSRG equations for disordered 1D bosons is identified and shown to
support the existence of long-range order on the percolation backbone
in two dimensions.  These results are relevant to experiments on
bosons in optical lattices and superconducting arrays, and also
(qualitatively) for the diluted Heisenberg antiferromagnet
La$_2$(Zn,Mg)$_x$Cu$_{1-x}$O$_4$.
\end{abstract}
\maketitle

\section{Introduction}
The physics of classical lattice percolation has been used to model a
great variety of physical systems with considerable
success,\cite{aharony} and recent results on 2D percolation have put
this classical theory on a firm mathematical footing.\cite{smirnov}
Several materials of current interest are well described by combining
random dilution with lattice quantum degrees of freedom, as in a
magnetic material in which some quantum spins have been removed by
chemical dilution.\cite{vajk} Aside from its interest as a microscopic
description of dilution in materials, percolation is important as the
simplest nondeterministic process for generating a ``fractal'', an
object of fractional dimensionality:\cite{mandelbrot} as explained
later in this introduction, there is a geometric phase transition in
randomly diluted lattice systems with a fractal structure at the
transition point.

The focus of this paper is on models for randomly diluted interacting
bosons.  These models describe superconducting Josephson-junction
arrays with only onsite interactions and a finite density of defective
junctions, or bosons in optical lattices where some sites remain
empty.  Percolation of superconductivity at short length scales has
also been discussed as a model for inhomogeneity observed by scanning
tunneling microscopy on the high-temperature superconductor
BSCCO.~\cite{lang} Finally, the $O(2)$ rotor is qualitatively similar
to the nearest-neighbor $s=1/2$ antiferromagnetic Heisenberg model in
that both have long-range order in two dimensions but only algebraic
correlations in one dimension.  The diluted Heisenberg model has been
studied numerically~\cite{vajk,sandvik,rong} and in neutron scattering
experiments on La$_2$(Zn,Mg)$_x$Cu$_{1-x}$O$_4$~\cite{vajk}.

Our interest in interacting bosons suggests that we consider the
diluted $O(2)$ quantum rotor model.  More generally, the $O(N)$
quantum rotor model \cite{sachdev} has an $N$ component unit vector
$\hat{n}_i$ living at every lattice site $i$.  The effect of dilution
is incorporated by defining the symbol $\alpha_{ij}$ to be unity if
the bond $ij$ is present in the diluted system, and zero otherwise.
For concreteness, we consider bond dilution though the results are
more general.  Then, the Hamiltonian for a particular realization of
the dilution $\{\alpha_{ij}\}$ is given by:
\begin{equation}
 H =\frac U2 \sum_i {\bf L}_i^2 - J\sum_{<ij>} \alpha_{ij} \hat{n}_i\cdot
\hat{n}_j 
\label{rotor}
\end{equation}
where, $U$ and $J$ are positive coupling constants, and the components
of the angular momenta $[{\bf L}]_{\alpha \beta}$ are given by $[{\bf
L}]_{\alpha \beta}=p_\alpha n_\beta - p_\beta n_\alpha$ ($\alpha \ne
\beta$) where $[n_\alpha, p_\beta]= i \delta_{\alpha \beta}$, and the
greek indices take values from $1$ to $N$. Thus the kinetic energy
part of the Hamiltonian is just ${\bf L}^2 = \sum_{\alpha<\beta}
L_{\alpha \beta}^2$.  For undiluted lattices of dimension $d\geq 2$,
the quantum $O(N)$ rotor model has long-range order, as long as
quantum fluctuations (measured by the ratio $U/J$) are below some
non-zero, $N$-dependent, critical value $(U/J)_c.$ However, a chain of
quantum rotors has no long-range order for any $N\geq 2$ unless
$U=0$.\cite{sachdev}

In a certain sense, a planar material with a large fraction of diluted
bonds (or sites) has geometry between one- and two-dimensions.  When
the dilution fraction $p$ reaches a lattice-dependent critical value
($p_c = 1/2$ for square lattice bond dilution), there is a geometric
phase transition: for $p<p_c$ there is an infinite nearest-neighbor
connected cluster of undiluted sites for essentially all realizations
of dilution, while for $p>p_c$ there is no infinite
cluster.\cite{aharony} At threshold ($p=p_c$), the number of bonds of
the cluster, $N(r),$ contained in a small circle of radius $r$ around
a given bond of the cluster: $N(r)\sim r^{d_f}$, with the ``fractal''
dimension $d_f=91/48\approx 1.896$[Ref. \onlinecite{aharony}].  Now,
an undiluted chain of sites has $N(r)\sim r$ and an undiluted square
lattice, $N(r)\sim r^2$, so, in this sense, the cluster is ``between''
one- and two-dimensions.  Note, however, that other geometric
properties of the cluster have other dimensionalities, some, in fact,
less than one.\cite{aharony}

Allowing quantum fluctuations ($U\neq 0$), raises an important
question about the nature of order in diluted, planar systems: Does
the order on the critical cluster which exists for $U=0$ persist for
non-zero $U$?  Early numerical and analytic work on the diluted
Heisenberg model suggested that long-range order vanishes before the
percolation threshold,~\cite{old_sandvik,castroneto} although recent
numerics disagree.~\cite{sandvik,castroneto2} This paper begins to
resolve this question analytically by coupling a spin-wave approach to
a complementary real-space renormalization group (RSRG) analysis.

The spin-wave approximation, described in Section II, provides a
natural first step for describing the effects of weak quantum
fluctuations on the $O(N)$ rotor model at threshold.  We show that the
{\it fracton} dimension $d_s$ of the cluster, which is close to $4/3$
for percolation in any dimension,\cite{aharony} is the relevant
dimension to consider when discussing ordering near percolation
threshold.  Since the fracton dimension is greater than one, quantum
fluctuations of the order parameter on an {\it average} site of the
cluster are small, according to our computation.  We also cite a
rigorous result that implies, within the framework of the spin-wave
approximation, that all sites have small quantum
fluctuations provided that $p<p_c.$ 

In principle, this is not enough to argue that the superfluid order is
stable to quantum fluctuations. One might worry whether the stability
to quantum fluctuations persists beyond the weak quantum fluctuations
allowed by the spin-wave approximation.  Further, at $p=p_c$ where the
spin-wave approach only describes the behavior of an {\it average}
site, one might be concerned that quantum fluctuations suppress
superfluid order on a set of sites of measure zero on the percolation
cluster. If these special sites are such that in their absence the
full connectivity of the cluster is lost, then there still may be no
true long range order. For instance the order may be destroyed on the
long 1D segments connecting the large blobs on the percolation cluster's
backbone (Fig. \ref{backbone}).

In Section III, we identify an effective model for rotors on the
cluster backbone that incorporates fluctuations along long links
beyond the linear approximation.  The physical idea, which we develop
in detail is that these fluctuations occur on a qualitatively faster
time-scale than fluctuations in the blobs.  We refer to this as the
slow blob approximation or SBA.  Within the SBA, we compute a
low-energy effective Hamiltonian for the $O(N)$ rotor model on the
percolation cluster.  Due to the one-dimensional topology of the
backbone, the effective Hamiltonian is a {\it one}-dimensional $O(N)$
rotor model with strongly-inhomogeneous charging $\{u_i\}$ and
exchange $\{j_i\}$ energies:
\begin{equation}
 H_{SB} = \sum_i \frac{u_i}{2} {\bf L}_i^2 - \sum_i j_i \hat{n}_i
  \cdot \hat{n}_{i+1},
\label{nolinks}
\end{equation}
where the coupling constants $\{j_i\},\{u_i\}$ are independent, random
variables, with distributions $P(j)$ and $R(u)$ respectively.  Like
the $\alpha_{ij}$ in (\ref{rotor}), these coupling constants are
randomly distributed.  However, the distributions $P(j),R(u)$ are
continuous for small $j\ll J$ and small $u\ll U$, respectively.  In
contrast the $\alpha_{ij}$ have a discontinuous, bimodal distribution.
Note that the SBA drastically simplifies the geometry of the problem.
In the SBA, the effective model (\ref{nolinks}) is truly
one-dimensional: namely the 1D $O(N)$ rotor model with strong
disorder.

To reach this effective description, we focus on the backbone of the
incipient infinite cluster.  Most bonds lie off the backbone: our
physical idea is that these ``dangling'' bonds will exhibit long-range
order if and only if the backbone bonds do so.  This is reasonable,
since at $U=0$, the dangling bonds play no role in communicating phase
correlations.  Moreover, previous work following this line of analysis
successfully explains a number of basic phenomena of ordering near
percolation threshold.  For example, the finite temperature phase
diagram in any dimension of {\it classical} magnets diluted with
non-magnetic impurities (e.g. $Rb_pMn_{1-p}F_2$ which has a large
$S=5/2$ local moment), follows from analyzing backbone
thermodynamics.\cite{aharony,coniglio_crossover,birgeneau} Further,
for quantum magnets with {\it discrete} symmetry (such as the
transverse field Ising model), backbone physics accounts for the zero
temperature phase diagram as well.\cite{senthil} The present work
simply extends the previous treatments to cover {\it continuous,
quantum} degrees of freedom.

Recently, the real-space renormalization group approach to quantum
systems with strong disorder, perfected by D.S. Fisher in the context
of random-exchange Heisenberg antiferromagnetic spin
chains,\cite{dsfisher} has been extended to the 1D $O(2)$ rotor model
with strong disorder by E.~Altman et.al.\cite{altman} In section IV
and V, the machinery of this approach is brought to bear on the
effective model in Eq.\ref{nolinks}.  The result indicates that for
the $O(2)$ rotor, superfluid order persists at threshold even with a
large amount of quantum fluctuation, namely, up to
$U/J\approx(U/J)_{KT}\approx\pi^2/4$, where, $(U/J)_{KT}$ is the
location of the Kosterlitz-Thouless transition of the clean 1D $O(2)$
rotor model.  Section V treats the range $U/J<(U/J)_{KT}$, while
section IV discusses the loss of superfluid order which occurs just
above $(U/J)_{KT}$.

Section VI concludes the paper with a discussion of the phase diagram
of the $O(2)$ rotor model with quenched dilution.  Two appendices
present details of the real-space renormalization group computations.

\section{Spin waves on the critical percolation cluster}

There are two main results of this analysis: First, exactly at
threshold ($p=p_c$) and except for certain rare points (``geometric
fluctuations''), long-range order is stable to weak quantum
fluctuations.  Second, away from threshold, on the percolating side
($p<p_c$), long-range order is stable to weak quantum fluctuations on
{\it any} site, for almost all clusters.

Clearly, if $U=0$ in (\ref{rotor}) we are in the classical limit of
this model, and the ground state simply has $\hat{n}_i=const.$ on all
sites of a connected cluster. The question that we will address first
is, exactly at the percolation threshold, does this long range order on
the percolation cluster survive the addition of small quantum
fluctuations?  We will address this question within a spin wave
calculation.  Since the spin wave calculation is essentially identical
for all $N \ge 2$, we specialize below to the O(2) model:
\begin{equation} 
H = \frac U2 \sum_{i} n_i^2
- J\sum_{< ij >}\alpha_{ij} \cos(\phi_i - \phi_j). 
\label{jja}
\end{equation}
 The $\phi_i$ represent the phase of the bosons at site $i$.  The
operator $n_i$ has integer eigenvalues and physically represents the
excess boson number at each site.  The $\phi_i$ and $n_i$ on a site
are conjugate variables:
\begin{equation}
[n_i,\exp i \phi_j] = \delta_{ij} \exp i\phi_i.
\end{equation}
This model may also be taken to represent an array of Josephson
junctions.  $U$ is a measure of the charging energy that induces
quantum fluctuations of the phase, $J $ is the strength of the
Josephson coupling.

In the absence of quantum fluctuations ($U=0$), the classical ground state is simply $\phi_i = const$. The effect of turning on a small $U/J$
may be addressed in a  harmonic (``spin-wave'') approximation
by simply expanding the cosine:
\begin{equation}
H_{sw} = \frac U2 \sum_{i} n_i^2 + \frac J2\sum_{< ij >}\alpha_{ij} (\phi_i - \phi_j)^2
\end{equation}

It is easier to work with the equivalent Euclidean (imaginary time) action:
\begin{eqnarray}
S &=& \int d\tau \sum_{i} \frac1{2U}(\partial_\tau \phi_i)^2 + \frac J2  \sum_{< ij >} \phi_i T_{ij} \phi_j \nonumber \\
  &=& \frac12 \int \bar{d}\omega (\frac{\omega^2}U\delta_{ij}+JT_{ij})\phi_i(\omega)\phi_j(-\omega)
\label{action}
\end{eqnarray}
where we have rewritten the potential energy part using $T_{ij} =
 \alpha_{ij}(2\delta_{ij}-1)$, and the frequency integral has
$\bar{d}\omega = \frac{d\omega}{2\pi}$ .

In this quadratic approximation, the superfluid order parameter is readily calculated:
\begin{eqnarray}
< \exp i \phi_i> &=&  \exp(-\frac12< \phi^2_i>) \nonumber \\
< \phi^2_i> &=& \int \bar{d}\omega \left[\left(\frac{\omega^2{\bf
1}}U + J{\bf T}\right)^{-1}\right]_{ii}  \\
 &  = & \sqrt{\frac{U}{J}}[{\bf T}^{-\frac12}]_{ii} \label{phisq}
\end{eqnarray}

Note that $<\phi_i^2>$ is formally of order $\sqrt{U/J}$. For
small $U/J$, it will therefore be small so long as $[{\bf
T}^{-\frac12}]_{ii}$ is finite, and the long range order will
survive quantum fluctuations. Below we will establish the required
finiteness.

Eqn. \ref{phisq} may be usefully rewritten in terms of the
eigenvalues and eigenvectors of the matrix ${\bf T}$. Since ${\bf
T}$ is a real, symmetric  positive semi-definite (${\bf
x^T T x}\ge 0$) matrix, it has real eigenvectors and eigenvalues
(${\bf \chi_n}, \lambda_n$) with $\lambda_n \ge 0$ (the zero
eigenvalue is obtained for the uniform vector on connected sites):
\begin{equation} 
\sum_j T_{ij}\chi_n(j) = \lambda_n \chi_n(i) 
\end{equation} 
Clearly, this
is just the eigenvalue problem for the Laplacian on the lattice
defined by the $\alpha_{ij}$, which in this case is taken to be at
the percolation threshold.
We get 
 \begin{equation} 
< \phi^2_i >  = \sum_n \frac1{\sqrt{\lambda_n}}[\chi_n(i)]^2
\label{phisq2} 
\end{equation}

The eigenvectors are assumed to be normalized on the percolation
cluster: \begin{equation}
\sum_{j \in perc. cluster} \chi_n^2(j) = 1
\label{norm}
\end{equation}

To argue for the finiteness of  the integral in
Eqn. (\ref{phisq2}), we first consider the site averaged value of the RMS phase
fluctuation:
\begin{eqnarray}
< \phi^2>_{Perc} &=& \frac1{N_p} \sum_{i\in Perc.cluster} <\phi_i^2>\\
 &=& \frac{1}{N_p}\sum_n \frac1 {\sqrt{\lambda_n}}
\end{eqnarray}

Here $N_p$ is the total number of sites on the incipient cluster
and we have used Eq.~(\ref{phisq2}), (\ref{norm}). In
terms of the density of states per site $\rho(\lambda)$ for the
eigenvalues of the operator ${\bf T}$

\begin{equation}
\rho(\lambda) = \frac{1}{N_p} \sum_n \delta(\lambda -
\lambda_n)
\end{equation}

we have: \begin{equation} < \phi^2>_{Perc} = \int_0^\infty d\lambda
\frac{\rho(\lambda)}{\sqrt{\lambda}} \end{equation}

Finiteness (or lack thereof) of the phase fluctuations is determined by the behaviour of the density of states at
very small $\lambda$. In the thermodynamic limit,
the asymptotic behaviour of $\rho(\lambda)$ is known
\cite{aharony}, since ${\bf T}$ is proportional to the Laplacian
on the percolation cluster. We have
\begin{equation} 
\rho(\lambda) \sim \lambda ^{\frac{d_s}2 - 1} 
\label{dos}
\end{equation}
where $d_s$ is called the {\it fracton} dimension. (On a regular
one dimensional line, we note that $d_s=1$, which gives the well known
logarithmic divergence in the RMS value of the phase
fluctuations). For percolation clusters in spatial dimension $d = 2$,
$d_s=1.32$ (see Ref.\cite{aharony}). In fact for percolation clusters
in all dimensions, its value is very close to $4/3$.

Thus, for percolation in $d=2$ we have:
\begin{equation} 
< \phi^2>_{Perc} \sim\int d\lambda \lambda^{-0.84}
\label{phisq4}
\end{equation} 
which is clearly finite when integrated over small $\lambda$.  Since
the fracton dimension of a percolation cluster does not vary much with
spatial dimension, this finiteness of the site averaged RMS phase
fluctuations holds for percolation in {\it any} dimension $d\geq 2$.

To reinforce this point, we turn to an alternate representation of the
suppression of superfluid order in terms of the return probability of
a random walker.  Using a recent rigorous result for the latter, we
can establish bounds on the former for an {\it any} site of the
cluster, provided that we are away from threshold on the percolating
side ($p<p_c$).  

Consider a random walker placed on the percolation cluster,
that begins its walk at site $i$. The differential equation describing
the time evolution of the probability of this walker being found at
time $t$ at position $j$ ($P_i(j,t)$) is:
\begin{equation}
\partial_t P_i(j,t) = -T_{jk} P_i(k,t)
\label{pev}
\end{equation} 
where an appropriate choice of the stepping rate has been made. Thus,
the probability of the particle returning to the site $i$ where it
started from is simply related to the eigenvalues and eigenfunctions
of the Laplacian operator on the percolation cluster:
\begin{equation} 
P_i(i,t) = \sum_n \chi_n^2(i)\exp (-\lambda_n t) 
\end{equation}
The root mean square (RMS) phase fluctuation amplitude at a site
$i$ (\ref{phisq2}) can be simply expressed in terms of this return
probability $P_{ii}(t)$ as 
\begin{equation} < \phi^2_i>  = \int_0^\infty dt
\frac{P_{i}(i,t)}{\sqrt{\pi t}} 
\label{phisq3} 
\end{equation}
As a simple check of this formula, consider the case of a linear
lattice, uniform, nearest-neighbor hopping
$J_{ij}=J_{i,i+1}=J$.  Then, the walker just performs a random walk in
one dimension.  If one takes the naive continuum limit of (\ref{pev}),
it is not hard to see that the return probability decays like
$1/\sqrt{t}$ as $t\rightarrow\infty$.  Applying (\ref{phisq3}), this
implies the usual logarithmic divergence in the RMS phase fluctuations
that destroys LRO in one dimensional superfluids \cite{hohenberg}.

For some percolation clusters, Remy and Mathieu have found an upper
bound for the asymptotic decay of the return probability.\cite{remy}
They considered site-percolation on the square lattice and looked at
strictly $p<p_c$: the random walker has an infinite connected cluster
to explore.  They showed rigorously that for almost
all cluster realizations, as $t\rightarrow\infty$,
\begin{equation}
P_i(i,t) < 1/t,,
\label{remy}
\end{equation}
where, $i$ refers to {\it any} site of the cluster.  In other words,
the asymptotic decay of the heat kernel on all sites of
almost all cluster realizations, decays as fast as it does in the
undiluted two-dimensional lattice.  Using (\ref{phisq3}), this implies
that for {\it any} site on the infinite cluster for $p<p_c$, the contribution
to RMS phase fluctuations from asymptotically long time-scales is
indeed finite, and bounded.  

The two main results of this section suggest that long-range order has
stability to weak quantum fluctuations.  We established these results
by making the spin-wave approximation to linearize the equations of
motion.  In the following section we introduce a complementary
approach (the slow-blob approximation) that treats the non-linear
dynamics beyond the spin-wave approximation.

\section{Slow blob approximation to O(N) model at $p=p_c$}

\begin{figure}
\begin{center}
\includegraphics[width=2.5in]{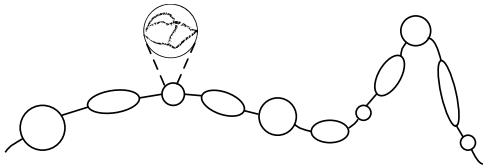}
\end{center}
\caption{The backbone of a percolation cluster is the union of
self-avoiding walks on the cluster.  It consists of blobs (ovals) connected by links.  The internal structure of a blob (inset) shows
that blobs consist of bonds which may be removed without disconnecting the cluster.  In contrast, links consist of bonds which, if removed, disconnect the cluster.}
\label{backbone}
\end{figure}

Within the slow-blob approximation to the backbone, we proceed in two
stages.  In the first stage, we treat blob degrees of freedom as
parameters rather than dynamical variables, and we solve the link
problem for the ground-state energy $E(\{\vec{L}_i,\vec{n}_i\})$.
Since links connect only to blobs (Fig. \ref{backbone}), the link
problem reduces to that of independent links.  In the second stage, we
treat the ground-state energy $E(\{\vec{L}_i,\vec{n}_i\})$ as an
adiabatic potential energy for the blobs, and solve for the blob
ground-state.

An analogous procedure arises in the study of molecular vibration
spectra, where it is called the Born-Oppenheimer approximation.\cite{born}  In
that case, the slow degrees of freedom, analogous to blobs, are the
massive ion cores.  The fast degrees of freedom, analogous to links,
are outer shell electrons.  Corrections to the Born-Oppenheimer
approximation are suppressed by the ratio $m/M\approx 10^{-3}$, of electron to
ion mass.

Here, the relevant control parameter is the ratio $u/U$ of blob
charging energy to bare charging energy.  If the blobs are internally ordered
then one has the simple relation
\begin{equation}
u=\frac{U}{n},
\label{blobcharge}
\end{equation}
where, $n$ is the number of sites in a given blob.  Now, on average,
blobs are very big: $<n>\approx L^{d_b-1/\nu}$, where $L$ is the
side-length of finite lattice.\cite{herrmann} See Table
\ref{tab:alpha}, for estimates of $\nu$ and $d_b$.  The bottom line is
that, in two and three dimensions $d_b-1/\nu>0$.  For $d_b>1/\nu$, the
blobs attached to the ends of a given link, have, on average, a
thermodynamically small charging energy: $u\sim U/L^{d_b-1/\nu}$.
Further, even the smallest blobs must contain at least four sites, so,
even in the worst case the blob charging energy $u/U\approx.25$ is
reasonably small.

This computation treats the blobs as internally ordered, which is a
reliable approximation. If the blobs were not internally ordered, then the
site-averaged RMS phase fluctuations $<\phi^2>_{\rm perc}$ would not
be finite.  In section II, we showed in the spin-wave
approximation that the fluctuations are finite (See
(\ref{phisq4})). Thus the blobs must be internally ordered, and
our computation in the preceding paragraph can be trusted.

Following the idea of the slow-boson approximation, we consider the
link problem with the blobs treated as parameters and determine $P(j)$
for $j\ll J$.  Consider a given disorder realization of the
$\alpha_{ij}$ in the bare Hamiltonian (\ref{rotor}).  Between two
neigboring blobs, the link length $\ell$ will vary randomly.  In fact,
the lengths have an exponential distribution:\cite{coniglio,herrmann}
\begin{equation}
p(\ell)\sim \exp(-\ell/\ell_0),
\label{ldist}
\end{equation}
with, for example, $\ell_0\approx 2.7$ for square lattice
percolation.\cite{herrmann} If we impose twisted boundary conditions
with $\vec{n}_1 \neq \vec{n}_2$, the ground-state energy of the link
increases by an $\ell$-dependent amount:
\begin{equation}
E(\vec{n}_1,\vec{n}_2)\approx - j(\ell)\vec{n}_1\cdot\vec{n_2},
\label{exchange}
\end{equation}
where, the ``stiffness'' $j(\ell)$, depends on the link length, $\ell$
and, implicitly, on the parameters of the bare model $J,U$ in
((\ref{rotor}).  From this we can deduce the distribution of blob
exchange couplings $P(j)$ appearing in (\ref{nolinks}) by a simple
change of variables:
\begin{eqnarray}
  P(j) = p(\ell) \left|\frac{{\rm d} j(\ell)}{{\rm d}\ell }\right| ^{-1}
\label{changevar}
\end{eqnarray}
where, $\ell$ is the link length satisfying $j(\ell)=j$.  
 
It remains to compute the stiffness of a link.  Fortunately, Cardy has
computed $j(\ell)$.\cite{cardy} For $N=2$, there are two qualitatively
different cases to consider: large $U$ ($U>\pi^2 J/4$) and small U
($U<\pi^2 J/4$).  For large $U$, the link breaks into segments of length
$\xi$ with the property that rotors in different segments are
essentially uncorrelated.  In this ``short-range'' phase, the
stiffness drops off exponentially with link length:
$j(\ell)\sim\exp(-\ell/\xi)$.  In contrast, for small $U$, the stiffness
falls off only as $\ell^{-1}$: $j(\ell)\rightarrow k /\ell$.  For
$N\geq3$, there is only one case to consider, since for all values of
$U/J$, the link is in a short-range phase with exponentially small
stiffness.\cite{sachdev} To summarize, the stiffness depends on link
length as follows:
\begin{equation} 
j(\ell) \sim \begin{cases}k/\ell,& N=2, ({\rm large} U)\\ 
e^{-\ell/\xi},& N\geq 3, {\rm and} N=2, ({\rm small} U).
\end{cases}
\label{single}
\end{equation}
%
Plugging this into (\ref{changevar}) gives the link strength distribution $P(j)$
\begin{equation} 
P(j) \sim \begin{cases}e^{-k/\ell_0 j},& N=2, ({\rm large} U)\\ 
j^{\xi/\ell_0-1},& N\geq 3, {\rm and} N=2, ({\rm small} U).
\end{cases}
\label{jdist}
\end{equation}

The distribution $R(u)$ follows directly from the distribution of blob
sizes $n$.  For large $n$, the fraction of blobs with size $n$, scales
like $n^{-2+\alpha}$, with $0<\alpha<1$, given in terms of backbone
dimension, $d_b$ and correlation length exponent $\nu$ via
$\alpha=1-1/(d_b\nu)$.\cite{herrmann} See Table \ref{tab:alpha} for
estimates of $\alpha$.  Thus, we have determined the charging energy
distribution entering (\ref{nolinks}), using (\ref{blobcharge}) to
change variables from blob size $n$ to charging energy $u$:
\begin{equation}
R(u) \propto u^{-\alpha},
\label{udist}
\end{equation}
where, $u\ll U$ is the charging energy of a given blob.  We have now
completely specified the blob problem (\ref{nolinks}).
\begin{table}
\caption{\label{tab:alpha} Correlation-length $\nu$, backbone $d_b$,
and blob-size $\alpha$ exponents for percolation in various
dimensions $d$.  For approximate results, number in parentheses indicates standard-deviation of last digit.}
\begin{ruledtabular}
\begin{tabular}{cccc}
 d       &  2        & 3                       & $\ge$ 6 \footnotemark[1] \\
\hline
$\nu$    & 4/3 \footnotemark[2] & 1.12(1)\footnotemark[3] & 1/2  \\
$d_b$    & 1.6431(6)\footnotemark[3] & 1.74(2)\footnotemark[3] & 2    \\
$\alpha$ & 0.5435(2)     & 0.487(10)                    & 0    \\
\end{tabular}
\end{ruledtabular}
\footnotetext[1]{Mean-field theory works for $d\ge6$ and gives exact
exponents, Ref.~\onlinecite{bunde}.}
\footnotetext[2]{Exact, see Ref.~\onlinecite{den_Nijs}}
\footnotetext[3]{Ref.~\onlinecite{herrmann}} 
\footnotetext[4]{Ref.~\onlinecite{zj}}
\end{table}
%


The goal of this model (\ref{nolinks}) is to understand the
competition between two possibilities: either very large blobs tend to
``anchor'' order across one-dimensional links, as has been suggested
to explain ordering in the diluted 2D Heisenberg model, or else
fluctuations within one-dimensional links destroy LRO.  Both
possibilities are realized in our model for different values of the
initial distributions.  For parameters chosen to reflect actual
percolation clusters in 2D, our renormalization group calculation,
described in section IV, finds LRO.

The next step is to see how this power-law
distribution of blob sizes enhances ordering and leads to new fixed
points in a real-space renormalization group calculation for
disordered interacting bosons in 1D.

\section{Renormalization group flow of $O(2)$ model for large $U$}

The idea of the real-space renormalization group for the $O(2)$ rotor
in 1D is to successively integrate out either the largest Josephson
coupling or the largest charging energy (See Appendix A for a
description of the elementary renormalization group step and its
generalization to the $O(N)$ case).  This iterative procedure
generates flow equations for the distributions of charging energy and
Josephson coupling.  The capacitance distribution is given as a
function $f(\z)$ of the scaled variable $\z = {\Omega/ u} - 1$, where
$\Omega$ is an upper cutoff of energy: $\Omega = \max_i \{u_i,j_i\}$,
and the Josephson coupling distribution is given as a function
$g(\beta)$ of the scaled variable $\beta = \log(\Omega/ j)$.

The analysis of one-dimensional models begins with the RSRG flow
equations~\cite{altman} as a function of energy scale $\G$ for the
charging energy distribution $f(\z,\G)$ and Josephson coupling
distribution $g(\beta,\G)$:
\begin{eqnarray}
{\partial f \over \partial \Gamma} &=& g_0\int_0^\infty \int_0^\infty f(\z_1) f(\z_2)
\delta(\z_1+\z_2 + 1 - \z)\,d\z_1\,d\z_2
\cr
&+&(1 + \z){\partial f \over
\partial \z}
+ (f_0+1-g_0) f.\cr
{\partial g \over \partial \Gamma} &=&  f_0 \int_0^\infty \int_0^\infty\,d\beta_1\,d\beta_2\,g(\beta_1) g(\beta_2) \delta(\beta_1 + \beta_2 - \beta)\cr
&+&{\partial g \over \partial \beta} + g (g_0 - f_0).
\label{floweq}
\end{eqnarray}
Here $\G=\log(\Omega_I/\Omega)$ tracks the progress of the
renormalization flow ($\Omega_I$ is the largest coupling in
(\ref{nolinks}) before renormalization), $f_0=f(0)$ and $g_0 = g(0)$.
%
%
We will eventually discuss the behavior of the coupled $f$ and $g$
equations, but for now, consider $g_0$ as a constant parameter
in the $f$ flow equation.  

It is possible to get immediate insight into the phase diagram found
for exponential distributions~\cite{altman} by averaging both sides of
the flow equation.  The average $\langle \z \rangle = \int_0^\infty \z
f(\z)\,d\z$, if it exists as for the exponential distributions
considered in by Altamn et. al.\cite{altman}, evolves as
\begin{equation}
{d \over d\G} \langle \z \rangle =
g_0 (1 + \langle \z \rangle) + f_0 \langle \z \rangle - 1 - \langle \z
\rangle.
\label{meanflow}
\end{equation} 
In this form, we can understand the appearance of a transition at
$g_0 = 1$ for the class of distributions studied in~[\onlinecite{altman}]: for
that one-parameter family of distributions $f_a(\z) = a e^{-\z a}$, so
that $f_0 = a$, we have $\langle \z \rangle = 1/a$ so \begin{equation}
f_0 \langle \z \rangle - 1 - \langle \z \rangle = -1/a = -1/f_0.
\end{equation} The flow equation for the mean capacitance, if the
function $f$ is initially of the form $f_a$, is \begin{equation} {d
\over d\G} \langle \z \rangle = g_0 (1 + 1/a) - 1/a = g_0 + (g_0 -
1)/a.
\label{meanflow2}
\end{equation}

The replacement $\delta(\z_1+\z_2 + 1 - \z) \rightarrow \delta(\z_1 +
\z_2 - \z)$ made in~\cite{altman} corresponds to replacing $g_0 (1 +
\langle \z \rangle)$ by $g_0 \langle \z \rangle$ in (\ref{meanflow}).
With this replacement, an $f$ distribution initially in the
one-parameter family remains in this one-parameter family even for
nonzero $g_0$, and the average charging energy, $a$, flows as
follows:~\cite{altman}
\begin{equation} 
{d a \over d\G} = (1 - g_0) a. 
\end{equation} 
This equation is written in terms of $a$ rather than $f_0$ in order to
stress that it is {\it specific to exponential distributions}.  It is
now shown that power-law distributions have a different RG equation
with different qualitative behavior, leading to a significantly
different phase diagram.


There are exact fixed points of the RG flow equation that are not
contained in the family of exponential distributions that has been
studied previously.  Power-law distributions are needed both to
describe these fixed points and to model the fractal distribution of
blob sizes on the percolation cluster [See Eq.~\ref{udist}].  Setting
$g_0=0$, we obtain a one-parameter family of static solutions of
(\ref{floweq})
\begin{equation} 
f_b(\z)={b\over (1+\zeta)^{1+b}}.
\label{fpsoln}
\end{equation} 
These are well-behaved decreasing normalized distributions for all
$b>0$.  For $b>1$, the mean $\langle \z \rangle$ converges and
satisfies
\begin{equation} 
f_0 \langle \z \rangle - \langle \z \rangle - 1 = 0
\end{equation}
so that, as required, the mean is constant in time.  Hence there is a
one-parameter family of fixed points $f_b$ that lies outside the
one-parameter family $f_a$ considered in~[\onlinecite{altman}], which contains
no fixed points at finite $a$.  Now if the function $f$ is originally
of the form $f_b$, the mean $\langle \z \rangle$ is increasing if $g_0
> 0$, suggesting that the system can flow to large capacitance (small
charging energy) if $g_0 > 0$.

More explicitly, we simply evaluate the right-hand side of the flow
equation (\ref{floweq}) using our power-law distributions
$f_b(\zeta),$ and find that the value of the distribution at large
capacitance increases.  That is, the
distribution flows to large capacitance assuming that it starts out
with a power-law form.  For a power-law form, the integral appearing
in (\ref{floweq}) has the following simple, asymptotic behavior:
\begin{equation}
\int_0^\infty \int_0^\infty f_b(\z_1) f_b(\z_2)\delta(\z_1+\z_2 + 1 - \z)\,d\z_1\,d\z_2 \rightarrow 2 f_b(\zeta),
\label{asymptotics}
\end{equation}
for $\zeta \gg 1.$ Using this fact, and the fact that at $g_0=0$,
power-laws are fixed points of the flow, one finds that the
$f_b(\zeta)$ distributions flow as follows at large capacitance:
\begin{equation}
{\partial f_b \over \partial \Gamma} \rightarrow g_0 f_b(\zeta).
\label{largezeta}
\end{equation}
For $g_0>0$, the right-hand side is positive, and, as the result, the
value of the distribution at large values of $\zeta$ increases under
the flow.  Thus, we see explicitly the tendency for power-law
distributions to flow towards large capacitance (small charging
energy).

So far we have seen that the mean $\langle \zeta \rangle$, assuming
$g_0 = 0$, is constant for the family $f_b$ and decreasing for the
family $f_a$.  There are also well-behaved distributions for which
$f_0 \langle \z \rangle - \langle \z \rangle - 1 > 0$: an example is
$f(\z) = 10 (1+\sqrt{\z})^{-6}$, for which $f_0 = 10$ and $\langle \z
\rangle$ = 1.  However, one might wonder if the initial increase or
decrease of the mean $\langle \z \rangle$ is a useful guide to the
eventual behavior of the RG flow.  To answer this question, Appendix B
constructs additional soliton solutions, which include the fixed
points $f_b(\zeta)$ as a limiting case, and for which the $g_0 = 0$
flow equation is exactly solvable.  One family of solutions are
related by a logarithmic rescaling of variables to solutions discussed
by Fisher in the context of spin chains.~\cite{dsfisher}  The
construction exhibits an interesting cumulant property of the
nonlinear flow equation at $g_0 = 0.$

The previous paragraphs constructed a new set of fixed-point
solutions, the $f_b(\z)$ in (\ref{fpsoln}), to the functional RG
equation with $g_0 = 0$.  Appendix B gives other solutions to the $g_0
= 0$ equation for $f$, but no full solution to the coupled RG
equations with $f_0$ and $g_0$ both nonzero has been found.  Hence
some other means of projecting the infinite-dimensional
renormalization group flows to a finite-dimensional subspace is
necessary.  This section gives a prescription for this projection that
both justifies the phase diagram found in [\onlinecite{altman}] for
exponential distributions and gives a new phase diagram once power-law
tails in capacitance are allowed.  The RG flows superficially depend
on a cutoff introduced in the projection process, but the phase
diagram and location of the critical point are cutoff-independent.

Let $0 < a < 1$ be some arbitrary cutoff, and define $0 \leq W_a
\leq 1$ as the integral 
\begin{equation} 
W_a = \int_a^\infty f(\z)\,d\z.
\end{equation} 
The evolution equation for $W_a$ is
\begin{equation} 
{d W_a \over d\G} = f_0 W_a - (1+a) f(a) + g_0
(1-W_a).
\label{waevol}
\end{equation} 
Note that for $g_0 = 0$, the fixed-point solutions $f_b(\z)$ have
$W_a = 1/(1+a)^b$ so that $dW_a / d\G = 0$ correctly.  In a moment we
will consider the flow generated by this equation within the $f_b$,
but first consider the exponential solutions $f(\z) = f_0 \exp(-f_0
\z)$.  Substituting into (\ref{waevol}) gives the projected flow 
\begin{equation}
{df_0 \over d\G} = f_0 - g_0 {e^{a f_0} - 1 \over a}.  
\end{equation}
 When $f_0$
is small, this equation becomes 
\begin{equation} 
{df_0 \over d\G} = f_0 - f_0 g_0,
\end{equation} 
which is exactly the RG flow obtained in~[\onlinecite{altman}] under an
approximation valid in this small-$f_0$ limit.  The same equation is
also obtained taking $a \rightarrow 0$ for any $f_0$.  For small
$f_0$, the superfluid transition in these equations at $g_0 = 1$ has
been discussed thoroughly by Altman et al.~\cite{altman}  The critical
point $f_0 = 0, g_0 = 1$ corresponds to the
transition between ``stiff'' and ``floppy'' regimes of the classical
XY model.~\cite{straley}  This RG flow is shown in Fig. 1.

Now consider instead the flow within the power-law solutions $f_b$.
For $g_0 \not = 0$, the evolution equation (\ref{waevol}) defines
a projected flow within the one-parameter subspace $f_b$: since $f_0 =
f(0) = b$, we have 
\begin{equation}
{d f_0 \over d\G} = -g_0 {1-W_a \over W_a \log
(1+a)} = - g_0 {\exp(f_0 \log (1+a)) - 1 \over \log(1+a)}.  
\end{equation}
Again taking the limit of small $f_0$ and any $a$, or small $a$ and
any $f_0$, we have 
\begin{equation} 
{d f_0 \over d\G} = -f_0 g_0.
\label{flowf}
\end{equation}
 Note that the right-hand-side is the same as in the
previously obtained flow equation for {\it coupling strength}~\cite{altman}
\begin{equation} 
{d g_0 \over d\G}
= -f_0 g_0 
\label{flowg}
\end{equation}
so that $f_0 - g_0$ is constant and the flows are as shown in Fig. 2.
This coupling strength flow equation describes the flow within a
power-law family of distribution of coupling strength $P(j)\propto
j^{g_0-1}$.  As argued in the previous section, such a distribution
enters (\ref{nolinks}) for the $O(2)$ rotor model with large $U$ [See
Eq.~\ref{jdist})].  In the next section, we discuss the small $U$ case.

Note that now there is no phase transition at the value $g_0 = 1$;
instead there are two lines of fixed points, one with $f_0 = 0$ and
one with $g_0 = 0$.  The entire classical line $f_0 = 0$ is stable,
including both the ``stiff'' and ``floppy'' regimes of the classical
XY model.  Now, we computed the initial conditions for this flow (See
(\ref{jdist}) and (\ref{udist})):
\begin{eqnarray}
f_0(\G=0)&=&1-\alpha, \nonumber \\
g_0(\G=0)&=&\xi/\ell_0,
\label{ic}
\end{eqnarray}
where, $\xi$ is the correlation length of a $O(2)$ rotor chain with
large charging energy $U,$ $\ell_0$ is the average length of links of red
sites, and, $\alpha=1-1/(d_b\nu)$ is the blob exponent tabulated in
Table \ref{tab:alpha}.  Since $f_0-g_0$ is a renormalization group
invariant of the flow in (\ref{flowf},\ref{flowg}), we can determine
the fixed point towards which the flow tends in a simple way.  Namely,
the flow tends towards the ordered, classical line $f_0=0$ provided
that initially we have $g_0(\G=0) > f_0(\G=0)$.  That is, to have
long-range order, we must tune $U/J$ close enough to the
Kosterlitz-Thouless transition of a {\it one-dimensional, clean}
$O(2)$ rotor model at $U/J\approx \pi^2/4$, where $\xi$ diverges, such that the
following criterion is satisfied:
\begin{equation}
\xi > \ell_0 (1-\alpha)
\label{criterion}
\end{equation}
Further, when this criterion is not satisfied, the flow tends towards
the insulating line of fixed points $g_0=0$.  Thus, the blob
hamiltonian (\ref{nolinks}) admits both insulating and superfluid
ground-states.  When $U/J$ is small enough to satisfy criterion
(\ref{criterion}), the renormalization group flow tends to the
superfluid ground-state.  The possibility of an ordered phase stable to
nonzero charging energy is novel, and supports the existence of order
for the $O(2)$ rotor model on percolation clusters via the connection
discussed in the Introduction.

\begin{figure}
\begin{center}
\includegraphics[width=2.5in]{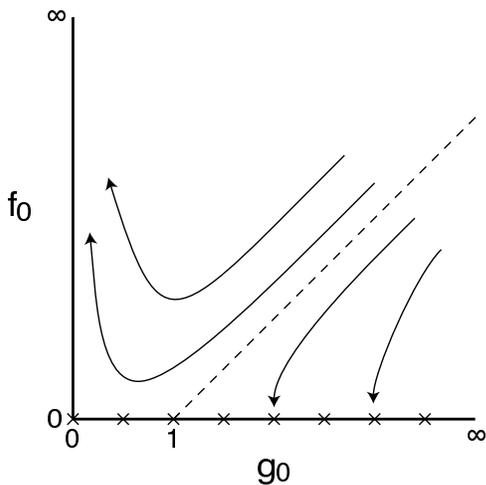}
\end{center}
\caption{Renormalization-group flows projected to two-parameter plane,
after~[\onlinecite{altman}], for capacitance distributions with exponential
damping at large capacitance.}
\label{figrg1}
\end{figure}

\begin{figure}
\begin{center}
\includegraphics[width=2.5in]{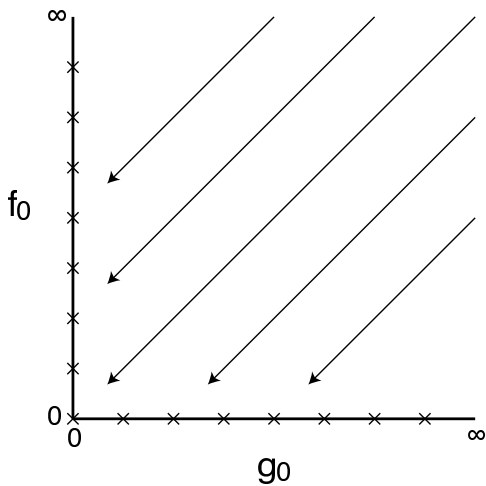}
\end{center}
\caption{Renormalization-group flows projected to two-parameter plane,
for capacitance distributions with power-law damping at large
capacitance.  Note the new line of fixed points at $g_0 = 0$ and the
absence of a phase transition at $(g_0,f_0) = (1,0)$.}
\label{figrg2}
\end{figure}

\section{$O(2)$ Rotor for small $U$}

In this section, we consider the special case of the $O(2)$ rotor with
small $U$, at percolation threshold, that is, (\ref{rotor}) with
$N=2$.  We will argue that this case is only slightly different than
the large $U$ case considered in the previous section.  Namely, we
will show using a spin-wave argument and a real-space renormalization
group argument, that this case exhibits long-range order regardless of
the value $U/J$, so long as $U$ is small.  Recall that, by contrast, for
large $U$, one must satisfy the criterion (\ref{criterion}) for the
ground-state to exhibit long-range order.

As in the previous section, we begin with the slow-blob approximation
(\ref{nolinks}).  In this approximation, the novelty of the $O(2)$
rotor with small $U$ arises from the ``activated'' form of the exchange
distribution $P(j)$ in (\ref{jdist}):
\begin{equation}
P(j) \sim e^{-k/\ell_0 j}, 
\label{jdist2}
\end{equation}
In other words, the power-law charging energy distribution
(\ref{udist}) holds for this case too.  Only the exchange distribution
changes compared to the large $U$ case.  Notice that the new, activated
exchange distribution is significantly less broad than the power-law
distributions considered in the previous section.

Let us, to first approximation, neglect completely the spread in the
exchange distribution, for this activated form.  This is a drastic
approximation which we will improve upon below by considering the flow
of the distribution under real-space renormalization.  However, the
activated form (\ref{jdist2}) is quite narrow, and treating it as
completely concentrated at its mean-value will turn out to be a
consistent approximation.  In this approximation, we may use the
following exact result for the asymptotic decay of the (appropriately
defined) return probability of a random walker:
\begin{equation}
\langle P_i(i,t)\rangle\rightarrow
(\frac{1}{\sqrt{t}})^{\frac{1}{1-\alpha/2}}.
\label{exact}
\end{equation}
Alexander et. al. \cite{alexander} show that this result obtains for
uniform exchange distribution and power-law charging energy
distribution $R(u)\propto u^{-\alpha}$, with $0<\alpha<1$.  Recall,
that this range of $\alpha$ obtains for percolation in two and three
dimensions (See table \ref{tab:alpha}).  Using (\ref{phisq3}), which
applies to (\ref{nolinks}), the asymptotic result in (\ref{exact})
implies a finite RMS phase fluctuation of the blob phases $\phi_i$,
where, the blob's rotor variable in (\ref{nolinks}) is given by
$\vec{n}_i=(\cos(\phi_i),\sin(\phi_i)),$ for $N=2$ rotor.  Thus, in
the approximation of uniform exchange energy, spin-wave theory
predicts that (\ref{nolinks}) exhibits an ordered phase stabilized by
the small-$U$ part of the charging energy distribution.

Let us now consider the actual distribution (\ref{jdist2}) and see how
it flows under real-space renormalization.  As in the previous
section, we gain immediate insight into the flow by averaging
both sides of the flow equation (\ref{floweq}).  The average $\langle
\beta \rangle = \int_0^\infty \beta g(\beta)\,d\beta$ exists for
activated distributions (\ref{jdist2}) and evolves as follows:
\begin{equation}
{d \over d\G} \langle \beta \rangle = -1 +  (g_0 +f_0) \langle \beta \rangle.
\label{meanflow3}
\end{equation} 
Recall, that $\beta=\log(\Omega/j)$ is the scaled Josephson exchange
and $g(\beta)=P(j=\Omega e^{-\beta})\Omega e^{-\beta}$ is the
distribution of $\beta$, where, $P(j)$ is given by (\ref{jdist2}).
More directly, the activated ansatz (\ref{jdist2}) implies a
distribution of $\beta$ of the following form:
\begin{equation}
g(\beta, \Gamma) = g_0(\G) e^{-g_0(\G) e^{\beta} -\beta +g_0(\G)},
\label{bdist}
\end{equation}
Here, $\G=\log(\Omega_I/\Omega)$ is the flow parameter, where,
$\Omega_I=\rm{max}\{j_i,u_i\}$ is the largest coupling in (\ref{nolinks})
before any renormalization, and $\Omega\leq\Omega_I$ is the largest coupling at
the current stage of renormalization.  The parameter, $g_0(\G)$ {\it initially}
satisfies $g_0(\G =0)= k/\ell_0\Omega_I$, .  At later times, we can
evaluate how $g_0(\G)$ flows by first computing the expectation
appearing in (\ref{meanflow3}):
\begin{equation}
\langle \beta \rangle = e^{g_0(\G)} E_1(g_0(\G)).
\label{expect}
\end{equation}
Here, the exponential integral appears: $E_1(g_0) = \int_{g_0}^\infty
 {\rm d}\gamma e^{-\gamma}/\gamma $.  Substituting (\ref{expect}) into
(\ref{meanflow3}) gives the projected flow of $g_0(\G)$:
\begin{equation}
\frac{ \,d g_0}{\,d \G} = \frac{ -1 +(f_0 + g_0) e^{g_0} E_1(g_0)}{-1+g_0 e^{g_0} E_1(g_0)} g_0. 
\label{dflow}
\end{equation}
When $g_0$ is small, the flow simplifies to the following:
\begin{equation}
\frac{ \,d g_0}{\,d \G} = g_0 + f_0 g_0 \ln g_0 + ..., 
\label{dflow2}
\end{equation}
where, the omitted terms are higher order in $g_0$.  The right-hand
side of (\ref{dflow2}) is initially positive, so $g_0$ grows under
renormalization.  Recall, that $g_0$ is the weight of the distribution
at the {\it large} values of the exchange strength $j$.  The projected
flow equation (\ref{dflow2}) implies that our activated form
(\ref{jdist2}) flows towards a narrower distribution concentrated at
the largest values of $j$.  This justifies the spin-wave treatment at
the beginning of this section (Eq.~(\ref{exact}) and paragraph containing
it), which assumed a narrow distribution of $j$.

Thus, the conclusion suggested by spin-wave theory holds even when we
allow the exchange distribution to start with an activated form.
Since this is precisely the form of distribution satisfied by the
$O(2)$ rotor with $U/J < \pi^2/4$, the model must have finite RMS fluctuations
of the blob rotor variables $\vec{n}_i$.  In the slow-blob
approximation, this implies that the percolation cluster as a whole
exhibits long-range order for the $O(2)$ rotor model with $U/J < \pi^2 /4$.  



\section{Conclusion}

To conclude, the question of whether the O(N) ($N\ge2$) quantum rotor
model can have an ordered phase on the percolation cluster was studied
via a spin wave calculation and via a real-space renormalization group
calculation.  The result for these models may be summarized in the
following way which also tracks the details of the calculation.  We
associate with the incipient percolation cluster an effective spatial
dimensionality in order to discuss the question of ordering.  The
correct spatial dimensionality for this purpose is the {\it fracton}
dimension $d_s$, defined in equation (\ref{dos}) from the spectral
density of the Laplacian on the percolation cluster. This
dimensionality $d_s$ is distinct from the fractal dimension of the
percolation cluster, and happens to be numerically very close to $4/3$
for percolation in any dimension. The possibility of long range order
can now be determined by simply comparing this dimension with the
lower critical dimension arising from the Hohenberg-Mermin-Wagner
theorem.\cite{hohenberg} Thus, for the zero temperature quantum models
where $d_s+1 > 2$ we have the possibility of long range order for
percolation in any dimension, while at finite temperatures, these
models are always disordered since $d_s < 2$.

The presence of weak links compelled us to investigate further the
possibility of long range order.  We ``integrated'' out the weak
links, generating a low-energy description [See Eq.~(\ref{nolinks})],
amenable to real-space renormalization.  The renormalization group
flow [See Fig.~\ref{figrg2} and Eq.~(\ref{flowf}),(\ref{flowg})]
demonstrated a competition between two natural length-scales
associated with the links [See Eq.~(\ref{criterion})].  On the one
hand, the average length of a link $\ell_0$, and on the other, the
correlation length within a long link $\xi$.  The final result may be
stated simply: When $U/J$ is small enough that $\xi >\ell_0
(1-\alpha)$, then correlations ``jump'' across the weak links and
create long-range order.  The renormalization-group flow supporting
this conclusion depends crucially on the fractal structure of the
percolation backbone.

\begin{figure}
\begin{center}
\includegraphics[width=2.8in]{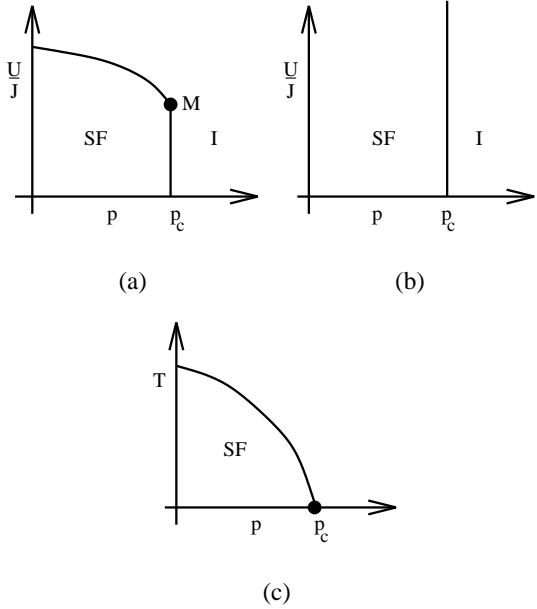}
\end{center}
\caption{The phase diagram for bosons with onsite repulsion on a
diluted lattice as a function of the dilution fraction $p$.  The
percolation transition occurs at $p_c$. (a) Integer filling at zero
temperature: The Mott insulating phase arises when charging energy $U$
is large compared to Josephson coupling $J$.  (b) Away from integer
filling, the Mott phase is not present for any $U/J$. (c) Finite
temperature transition for bosons out of the superfluid phase.}
\label{phasediag}
\end{figure}

These considerations are summarized in the phase diagrams in
Fig.~\ref{phasediag}.  In Fig.~\ref{phasediag}(a), the existence of a
finite superfluid order parameter at the percolation transition for
small $U/J$ leads to the vertical phase boundary.  In
Fig.~\ref{phasediag}(b), we simply point out that the well-known
absence of a Mott insulating state for incommensurate filling at zero
dilution persists under dilution until percolation threshold.  Finally,
the absence of superfluid order at finite temperatures on percolation
clusters in any dimension $D\ge 2$ gives rise to the phase diagram
shown in Fig.~\ref{phasediag}(c).

\section*{Acknowledgments}

This project was inspired by scintillating discussions with Antonio
Castro Neto, G. Refael, O. Motrunich, E. Altman, E. Mucciolo and
C. Chamon, for which we are grateful.  We thank E. Mucciolo and
C. Chamon for also sharing their results prior to publication. This
work was supported by NSF DMR-0238760 and the Hellman Foundation
(N. B.-A. and J. E. M.), a Pappalardo Fellowship (AV) and by the MRSEC
program of the NSF under DMR-0213282 (TS).  TS also acknowledges
support from the Alfred P. Sloan Foundation, The Research Corporation,
and NEC corporation.

\appendix

\section{RG of random O(N) chain}
Consider the $O(N)$ rotor chain with random exchange and ``charging''
given in Eq.~({\ref{nolinks}).  We have two possible rg steps.  First, if
the largest energy is an exchange $J_i$, then we lock the two rotors
together, and estimate the effective charging energy of the combined
rotor.  The result is that inverse charging energies add:
\begin{equation}
\tilde{U}^{-1}\approx U_i^{-1}+U_{i+1}^{-1}
\label{ustep}
\end{equation}
Second, the largest energy may be a charging energy $\Omega=U_i$.  In this
case, we put this rotor in the rotationally invariant ground-state,
and treat the neighboring rotors as a classical magnetic field
$\vec{h}=J_i \vec{n}_i+J_{i+1}\vec{n}_{i+2}$.  For small
$|\vec{h}|/\Omega$, the ground-state energy is, to lowest order, $2h^2/N(N-1)\Omega$.
Expanding this, yields a new effective coupling between the
neighboring rotors:
\begin{equation}
\tilde{J}\approx\frac{4J_i J_{i+1}}{N(N-1)\Omega}
\label{jstep}
\end{equation}
The first step is identical to O(2) case.  The second step is
virtually identical.  If we measure exchanges on a logarithmic scale,
$\beta_i=\log(\Omega/J_i)$, then the $\Omega=U$ step becomes
\begin{equation}
\tilde{\beta}=\beta_i + \beta_{i+1} + \log(N(N-1)/4).
\label{jstep2}
\end{equation}
As always, in asymptotic analysis of these kind of flow equations, we
can neglect additive constants in the RG-step, since we are looking
for broadly distributed fixed points.\cite{dsfisher} In conclusion, we
have obtained the elementary RG-step for the $O(N)$ rotor chain with
strong disorder.

\section{Additional solutions of the RSRG flow equation}

The goal of this section is to find $\G$-dependent exact solutions of
the capacitance flow equation (\ref{floweq}) that verify the
statements in Section III derived using projected RG flows.  Although
we have not been able to solve the nonlinear PDEs for arbitrary
initial data, the exact solutions now discussed are examples
supporting the qualitative behavior found using a moment equation in
Section III.  Let us look for a solution in terms of a distribution
$\phi(b,\G)$ over the one-parameter family $f_b(\z)$:
\begin{equation} 
f(\z,\G) = \int_0^\infty
\phi(b,\G) f_b(\z)\,db = \int_0^\infty \phi(b,\G) {b \over
(1+\z)^{1+b}}\,db.  
\end{equation} 
Now $f_0(\G) = f(0,\G) = \int_0^\infty b
\phi(b,\G)\,db$, i.e., the average $\langle b \rangle_\phi$ taken over
the distribution $\phi(b,\G)$.  Our goal will be to find solutions of
the evolution equation for $\phi(b,\G)$: 
\begin{equation}
 {\partial \phi(b,\G)
\over \partial \G} = \left(-b + \int_0^\infty \phi(b_2) b_2\,db_2
\right) \phi(b,\G).  
\end{equation} 
Now assume that $\phi(b,\G)$ is nonnegative
(it is then automatically normalized to 1) so that it can be regarded
as a probability distribution.  Its normalization is constant, and its
mean (equal to $f_0$) evolves via 
\begin{equation} 
{d \langle b \rangle_\phi \over
d\G} = {\langle b \rangle_\phi}^2 - \langle b^2 \rangle_\phi.  
\end{equation} 
In terms of the mean $\mu_1$ and the variance $\mu_2$, this equation
is compactly written ${d \mu_1 \over d\G} = - \mu_2$.  A bit of
algebra confirms that a similar evolution holds for all cumulants:
\begin{equation} 
{d \mu_n \over d\G} = - \mu_{n+1}.  
\end{equation}
This provides a way to find soliton-like solutions.  One class of
solutions starts from the generalized Poisson distribution with two
parameters $\lambda$ and $x_0$, 
\begin{equation} 
\phi(b) =
\sum_{k=0}^\infty \delta(b-k x_0) {\lambda^k e^{-\lambda} \over k!}
\label{poisson}
\end{equation} 
which has cumulants $\mu_n = \lambda {x_0}^n$.  This distribution
has a finite probability of $b=0$, which corresponds to exactly 0
charging energy; below we find a different soliton solution without
this property.  The discrete distribution (\ref{poisson}) approaches a
continuous Gaussian centered at $b=b_0$ if we take the limits $x_0
\rightarrow 0$ and $\lambda \rightarrow \infty$ with $\lambda x_0 =
b_0$.  Then the cumulant evolution equations are all solved if $x_0$
is constant and $\lambda(\G)$ solves 
\begin{equation} {d \lambda \over
\G} = - x_0 \lambda \Rightarrow \lambda(\G) = \lambda(0) e^{- x_0 \G}.
\end{equation} 
In the limit of small $x_0$ with $x_0 \lambda(0)$
fixed, we recover the fixed-point solutions found previously.

A second family of solutions starts from the gamma distribution with
parameters $\alpha$ and $\theta$: 
\begin{equation}
 \phi(b) = {b^{\alpha - 1}
e^{-b/\theta} \over \G(\alpha) \theta^\alpha}.  
\end{equation}
 The cumulants of
the above distribution are 
\begin{equation} 
\kappa_n = \G(n) \alpha \theta^n.
\end{equation}
So, if the only $\G$ dependence is in $\theta$, 
\begin{equation} {d\kappa_n
\over d\G} = \G(n+1) \alpha \theta^{n-1} {d\theta \over d\G}= - \alpha
\theta^{n+1} \G(n+1) = - \kappa_{n+1} 
\end{equation} 
if
\begin{equation} {d \theta \over d\G} = - \theta^2 \Rightarrow
\theta(\G) = {1 \over \G + \theta(0)^{-1}}.  
\end{equation} 
It is
simple to confirm explicitly that the resulting form for $f$,
\begin{equation} 
f(\z,\G) = {\alpha \theta(\G) \over (1+\z) (1+\theta
(\G) \log(1+\z))^{\alpha+1}}, 
\end{equation}
 is normalized for $\alpha
> 0$ and satisfies the evolution equation.  For this solution,
$f_0(\G) = \alpha/(\G+\theta(0)^{-1})$, indicating that this charging
energy distribution is marginally irrelevant even as $g_0 \rightarrow
0$, well away from the classical transition.

We also note that the gamma solutions have normalized variance
$\kappa_2 / (\kappa_1)^2 = 1/\alpha$, which is a constant of the
motion.  As $\alpha \rightarrow 0$, the distributions become extremely
broad, and the RG is expected to be valid in this limit.

These two families of solutions suggest that having a finite
probability of zero charging energy, as in the Poisson case, leads to
an exponentially rapid flow of the entire distribution to zero
charging energy: the order is strongly ``nucleated'' by the classical
rotors with zero charging energy.  The gamma solutions also flow to
zero charging energy, proving that such flow is possible for regular
distributions, but this solution flows only as a power-law.  Note that
the flow equation for $\theta$ is exactly that for a marginally
irrelevant operator.  


\bibliography{rgnotes_jan1}
\end{document}